\documentclass[12pt]{article}
\usepackage{graphicx}
\usepackage{epstopdf}
\usepackage{amsmath}
\usepackage{amsfonts}
\usepackage{amssymb}
\usepackage{color}
\usepackage{mathrsfs}

\pdfoutput=1

\newcommand{\be}{\begin{equation}}
\newcommand{\ee}{\end{equation}}
\newcommand{\bea}{\begin{eqnarray}}
\newcommand{\eea}{\end{eqnarray}}
\newcommand{\barr}{\begin{array}}
\newcommand{\earr}{\end{array}}

\newcommand{\hinvMpc}{\,h\, {\rm Mpc}^{-1}\,}

\def\km{{k_{\rm M}}}
\def\vk{{\vec k}}
\def\xperp{{\vec x}_\perp}
\def\qperp{{\vec q}_\perp}
\def\zperp{{\vec z}_\perp(\vec q,t)}
\def\sperp{{\vec s}_\perp(\vec q,t)}
\def\sperpo{{\vec s}_\perp(\vec 0,t)}

\def\kperp{{\vec k}_\perp}
\def\tkperp{{\vec {\tilde k}}_\perp}

\def\xparal{{x}_\parallel}
\def\qparal{{q}_\parallel}
\def\zparal{{z}_\parallel(\vec q,t)}
\def\sparal{{s}_\parallel(\vec q,t)}
\def\sparalo{{s}_\parallel(\vec 0,t)}
\def\dsparal{{\dot s}_\parallel(\vec q,t)}
\def\dsparalo{{\dot s}_\parallel(\vec 0,t)}
\def\kparal{{k}_\parallel}
\def\tkparal{{\tilde k}_\parallel}

\def\epssm{{\epsilon_{s<}}}
\def\epssM{{\epsilon_{s>}}}
\def\epsdm{{\epsilon_{\delta<}}}

\usepackage[colorlinks,bookmarks]{hyperref}
\definecolor{linkblue}{rgb}{0,0,0.8}
\definecolor{linkgreen}{rgb}{0,0.5,0}

\hypersetup{pdfpagemode=UseNone, pdfstartview=FitH, linkcolor=linkblue, %
            citecolor=linkgreen, urlcolor=linkblue}

\def\beq{\begin{equation}}
\def\eeq{\end{equation}}
\def\be{\begin{equation}}
\def\ee{\end{equation}}
\def\bea{\begin{eqnarray}}
\def\eea{\end{eqnarray}}
\def\d{{\partial}}

\def\nn{\nonumber}

\def\knl{{k_{\rm NL}}}

\begin{document}


\setcounter{page}{1} \baselineskip=15.5pt \thispagestyle{empty}

\begin{flushright}
\end{flushright}

\begin{center}

{\Large \bf Redshift Space Distortions\\[0.5cm] in the Effective Field Theory of Large Scale Structures}
\\[0.7cm]
{\large Leonardo Senatore${}^{1,2}$  and Matias Zaldarriaga${}^3$}
\\[0.7cm]
{\normalsize { \sl $^{1}$ Stanford Institute for Theoretical Physics,\\ Stanford University, Stanford, CA 94306}}\\
\vspace{.3cm}

{\normalsize { \sl $^{2}$ Kavli Institute for Particle Astrophysics and Cosmology, \\
Stanford University and SLAC, Menlo Park, CA 94025}}\\
\vspace{.3cm}

{\normalsize { \sl $^{3}$ School of Natural Sciences, Institute for Advanced Study, \\Olden Lane, 
Princeton, NJ 08540, USA}}\\
\vspace{.3cm}

\end{center}

\vspace{.8cm}

\hrule \vspace{0.3cm}
{\small  \noindent \textbf{Abstract} \\[0.3cm]
\noindent We introduce a formalism, valid both for dark matter and collapsed objects, that allows us to describe redshift space distortions in the context of the Effective Field Theory of Large Scale Structures (EFTofLSS). Expressing density perturbations in redshift space corresponds to performing a change of coordinates and the resulting expressions contain  products of density perturbations and velocity fields evaluated at the same location. These terms are sensitive to non-perturbative short-distance physics and in order to correctly treat them they need to be renormalized by adding suitable counterterms.  Therefore more counterterms are required in redshift space expressions compared to their real space analogs. In particular in the expression for the one-loop matter power spectrum there are two new counterterms. 
Just as in real space, long wavelength displacements affect correlation functions in redshift space and need to be  resummed. We generalize the real space  formulas for IR resummation to this case: the final expressions are conceptually similar but are more challenging to compute numerically due to their reduced symmetry.

 \vspace{0.3cm}
\hrule


\section{Introduction}

After the Planck satellite delivers its final results, obtaining additional cosmological information will probably depend on our ability to model the development of Large Scale Structure (LSS).  It is therefore crucial to reliably predicts the associated observables. On large scales, density perturbations  are small, while they are large on short scales, where gravitational collapse happened and halos and galaxies formed. One approach that is very common is to perform large  $N$-body simulations, that reproduce both the linear and the non-linear regimes. While this is an acceptable approach, it has some complications due to the long time it takes to run the simulations and scan over cosmological parameters and over new physical processes and effects that might be operating.  

On large scales, perturbations are small and one expects to be able to predict the evolution of the density perturbations in an analytic, and therefore rather simple, way. Perturbation theory for LSS has a long history dating back to the very early days of modern Cosmology {\it e.g.} \cite{Zeldovich:1969sb,Peebles:LSS}. It is extremely successful at calculating correlators at the lowest order or tree level (for a complete review of perturbation theory results see for example \cite{Bernardeau:2001qr}.) However results for the first nontrivial correction to tree level results, the ``loop corrections", have been less than satisfactory. 

Recently the introduction of the so-called Effective Field Theory of Large Scale Structures (EFTofLSS)~\cite{Baumann:2010tm,Carrasco:2012cv} has led to substantial progress. In this approach, the effect that short distance non-linear physics has on large distances is taken into account by adding suitable counterterms that are renormalized during the perturbative calculation to the correct physical values. In this way, the calculation for Large Scale Structure (LSS) observables is organized in a perturbative expansion in powers of the parameters $\epssM$, $\epsdm$ and $\epssm$, which are defined as~\cite{Porto:2013qua}
\bea
&&\epsilon_{s >} =k^2  \int_k^\infty {d^3k' \over (2 \pi)^3}  {P_{11}(k') \over k'^2}\ , \\ \nonumber
&&\epsilon_{\delta <} = \int^k_0 {d^3k' \over (2 \pi)^3} P_{11}(k')\ , \\ \nonumber
&&\epsilon_{s_<} =k^2  \int_0^k {d^3k' \over (2 \pi)^3}  {P_{11}(k') \over k'^2}\ ,
\eea
and respectively represent the effect at a given wavenumber $k$ of the displacement due to shorter wavelength perturbation, of the matter overdensities induced by longer wavelength perturbations, and finally of the displacement due to the longer wavelength perturbations.

It turns out that for many of the modes of interest in the current universe the parameter $\epsilon_{s_<}$ is of order one and therefore a perturbative treatment is not appropriate. While, as far as the authors of this paper know, there is no hope of resumming perturbations that are dominated by short distance contributions, one can do so for long wavelengths which are well described by the linear regime. This has allowed a reformulation of  the EFTofLSS that treats the parameter $\epssm$ non-perturbatively~\cite{Senatore:2014via}. The same approach has been recently extended to biased tracers in~\cite{Senatore:2014eva}.

When compared to numerical data for dark matter correlations, the EFTofLSS has performed remarkably well. At one loop, the matter power spectrum agrees with $N$-body simulations to percent level up to wavenumber $k\sim 0.3\hinvMpc$~\cite{Carrasco:2012cv}. Predictions depend on one free parameter,   the so-called speed of sound, which can be measured either by fitting to the power spectrum or directly in small $N$-body simulations~\cite{Carrasco:2012cv}. At one loop the momentum power spectrum~\cite{Senatore:2014via} and the bispectrum~\cite{Angulo:2014tfa, Baldauf:2014qfa} agree with similar accuracy to $N$-body data up to the same wavenumber $k\sim 0.3\hinvMpc$. These are very nice consistency checks of the EFTofLSS, as the EFT is expected to perform equally well on all correlation functions when they are computed at the same order. Another important consistency check passed by the EFTofLSS has been the prediction of the slope of the velocity vorticity power spectrum which matches the one measured in simulations~\cite{Carrasco:2013mua,Mercolli:2013bsa}. 

At two-loops the EFTofLSS agrees with the power spectrum of dark matter up to a very high wavenumber $k\sim 0.6\hinvMpc$~\cite{Senatore:2014via,Carrasco:2013mua}. This is a remarkable result because it suggests that we can have perturbative control on very small scales. If the same reach is maintained in all observables, the consequences  for next generation large scale structure surveys could be huge.  The number of available modes would increase by a large factor which would lead to very significant improvements for the capability of this experiments to constrain properties of neutrinos, dark energy, the primordial power spectrum, and, most importantly, primordial non-Gaussianities. 

The purpose of this paper is to develop the formalism that allows us to include the treatment of redshift space distortions within the EFTofLSS. We will do this both for dark matter and for collapsed objects, so-called biased tracers.

\section{Redshift Space Distortions in the EFTofLSS}

\subsection{Eulerian Description\label{sec:eulerian}}

In this section we describe how to introduce redshift space distortions in the context of the EFTofLSS. 
We start with the Eulerian treatment, where we expand in all perturbations treating them equally.  We will upgrade this treatment by resumming the IR-modes in the next section. 

For algebraic simplicity, we work in the `distant observer' approximation, where the line of sight direction is fixed in space and denoted by $\hat z$. We neglect all general relativistic effects as well as evolution effects between the various points we correlate. Since we are interested in improving the treatment of non-linear corrections to the clustering of LSS this last approximation is particularly justified: relativistic effects and evolution effects are important only on very large distances of the order of the Hubble scale, where linear theory suffices.

 The relation between the position in real space $\vec x$ and in redshift space $\vec x_r$ is given by (see for example~\cite{Matsubara:2007wj}):
\be
\vec x_r=\vec x+\frac{\hat z\cdot \vec v}{a H} \hat z\ .
\ee
Mass conservation relates the density in real space $\rho(\vec x)$ and in redshift space $\rho_r(\vec x_r)$:
\be
\rho_r(\vec x_r)\; d^3x_r=\rho(\vec x)\; d^3 x\ ,
\ee
which implies  
\be
\delta_r(\vec x_r)=\left[1+\delta\left(\vec x(\vec x_r)\right)\right] \left|\frac{\d\vec x_r}{\d \vec x}\right|_{\vec x(\vec x_r)}^{-1}-1\ .
\ee
In Fourier space, this relationship becomes 
\be\label{eq:first}
\delta_r(\vec k)=\delta(\vec k)+\int d^3 x \; e^{-i \vec k\cdot\vec x} \left(\exp\left[-i \frac{k_z }{a H} v_z(\vec x)\right]-1\right)\left(1+\delta(\vec x)\right)\ .
\ee
We now assume we can Taylor expand the exponential of the velocity field to obtain an expression that is more amenable to perturbation theory (this is where the Eulerian approach that we describe here, and the Lagrangian approach that we describe later differ). For the purpose of this paper, we will show formulas that are valid only up to one loop. We therefore can Taylor expand up to cubic order, to obtain
\bea\label{eq:delta_redshift_1}
&&\delta_r(\vec k)\simeq\delta(\vec k)+\\ \nonumber
&&\quad \qquad \int d^3 x \; e^{-i \vec k\cdot\vec x} \left[ \left(-i \frac{k_z}{aH} v_z(\vec x)+\frac{i^2}{2}\left(\frac{k_z}{aH}\right)^2 v_z(\vec x)^2-\frac{i^3}{3!}\left(\frac{k_z}{aH}\right)^3 v_z(\vec x)^3\right)\right.\\ \nonumber
&&\ \ \qquad\qquad \qquad\qquad\left.+ \left(-i \frac{k_z}{aH} v_z(\vec x)+\frac{i^2}{2}\left(\frac{k_z}{aH}\right)^2 v_z(\vec x)^2\right)\delta(\vec x) \right]\\ \nonumber
&&\quad =\delta(\vec k)-i \frac{k_z}{aH} v_z(\vec k)+\frac{i^2}{2}\left(\frac{k_z}{aH}\right)^2 [v_z^2]_\vk-\frac{i^3}{3!}\left(\frac{k_z}{aH}\right)^3 [v_z^3]_\vk-i \frac{k_z}{aH} [v_z \delta]_\vk+\frac{i^2}{2}\left(\frac{k_z}{aH}\right)^2 [v_z^2\delta]_\vk \ ,
\eea
where in the last line we have introduced the notation $[f]_\vk=\int d^3x\; e^{-i \vec k\cdot\vec x}\,f(\vec x)$.

Equation (\ref{eq:delta_redshift_1}) is problematic because it contains  products of operators at the same point in space. One encounters a similar situation in the EFTofLSS when computing correlators involving the velocity field~\cite{Carrasco:2013mua,Mercolli:2013bsa}. The velocity field is defined as the ratio of the momentum operator,~$\vec \pi(\vec x,t)$ and the density operator $\rho(\vec x,t)$:
\be\label{eq:velocity_def}
v^i(\vec x,t)=\frac{\pi^i(\vec x,t)}{\rho(\vec x,t)}\ .
\ee
 This creates a problem because in the EFTofLSS we admit that we have no detailed knowledge of what happens at distances shorter than the non-linear scale, and therefore there is no sense in which this ratio can be taken in a controlled way. 
 
Fortunately  we are not interested in describing the velocity field at very short distances, but we are interested in long-wavelength correlation functions.  As is standard in the EFTofLSS, our ignorance of the effects at long distances from short distance physics can be encapsulated in a few operators that have to be added to the  original one order by order in perturbation theory.  Coefficients need to be adjusted to renormalize the theory.  

 For example, for the velocity field, one has 
\be\label{eq:velocity_def_ren}
v_{R}^i(\vec x,t)=v^i(\vec x,t)-\int \frac{d a'}{a'} \frac{H(a')}{\knl^2} \,K_{\rm v}(a,a')\, \d^i\delta(a', \vec x_{\rm fl})+\ldots\ ,
\ee
where $K_V$ is a free function dubbed counterterm, $H$ is the Hubble rate, $\knl$ is the nonlinear scale, $a$ and $a'$ represent the scale factor, and the dots represent higher derivative and higher order terms. Finally, $\vec x_{\rm fl}$ is defined recursively as
\beq
\vec x_{\rm fl}[\tau,\tau'] = \vec x - \int_{\tau'}^{\tau} d\tau''\; \vec{v}(\tau'',\vec x_{\rm fl}[\tau,\tau''])\ .
\eeq
In the absence of counterterms, correlation functions of $\vec v$ receive uncontrolled and possibly even infinitely large contributions from short distance physics. $K_V$ and the additional higher derivative or higher order counterterms have the role of  removing this UV dependence in $\vec v$ so that correlation functions of $\vec v_{R}$ are finite and receive contribution only from modes up to $\knl$.  The integral in $a'$  is present because the EFTofLSS is local in space, but {\it non-}local in time~\cite{Carrasco:2013mua,Carroll:2013oxa}. For calculations up to one loop order, the time non-locality is degenerate with a local treatment, by a redefinition of the coefficients~\cite{Carrasco:2013mua}. We can therefore re-write~(\ref{eq:velocity_def_ren}) as
\be\label{eq:velocity_def_ren_loc}
v_{R}^i(\vec x,t)=v^i(\vec x,t)-c_{v,\delta}(t)\, \frac{H}{\knl} \frac{\d^i}{\knl}\delta(\vec x,t)+\ldots\ .
\ee
In our expression for the redshfit space distortion~(\ref{eq:delta_redshift_1}), we are faced with the same situation as for the definition of the velocity field and thus we will have to add appropriate counterterms. Before proceeding it is useful to re-write~(\ref{eq:delta_redshift_1}) in terms of the momentum as
\bea\label{eq:delta_redshift_2}
&&\delta_r(\vec k) =\delta(\vec k)-\frac{1}{\rho_b}i \frac{k_z}{aH} \pi_z(\vec k)+\frac{i^2}{2}\left(\frac{k_z}{aH}\right)^2 [v_z^2]_\vk-\frac{i^3}{3!}\left(\frac{k_z}{a H}\right)^3 [v_z^3]_\vk+\frac{i^2}{2}\left(\frac{k_z}{aH}\right)^2 [v_z^2\delta]_\vk \ ,
\eea
where $\rho_b$ is the background density.
This is useful because, as explained in~\cite{Carrasco:2013mua}, the momentum $\pi^i$ and $\delta$ are automatically renormalized by the same counterterms. We can write the momentum field as the sum of a scalar $\pi_S$ and a vector component $\vec \pi_V$, with $\d_i\pi_V{}^i=0$:
\be
\pi^i=a \rho_b\left[\frac{\d_i}{\d^2}\pi_S+\epsilon^{ijk} \frac{\d_j}{\d^2} \pi_{V,k}\right]\ ,
\ee
where $\pi_S$ starts at linear order while $\vec \pi_V$ starts at cubic order. We have chosen the factors of $a$ and $\rho_b$ so that, from the continuity equation, $\pi_S=-\dot\delta$. Notice that under a parity transformation $\pi_S$ is a even, while $\vec \pi_V$ is odd. The equation for the vector part of the momentum follow trivially from the momentum equation provided in~\cite{Carrasco:2013mua}, and is given by
\be
a\rho_b\left[\dot \pi_V{}^i+2 H \pi_V{}^i\right]-\epsilon^{ijk} \frac{\d_j\d_l}{a}\left(\frac{\pi^l\pi^k}{\rho_b(1+\delta)}\right)-\epsilon^{ijk} \frac{\d_j}{a}\left(\rho_b(1+\delta)\d_k\phi\right)=\epsilon^{ijk}\frac{\d_j\d_l}{a}\tau^{il}\ ,
\ee
where $\tau^{il}$ is the effective stress tensor. Since $\pi_V$ vanishes in the initial conditions, it can be checked it starts at cubic order. At the one-loop order that concerns us, the cubic order of $\pi_V$, $\pi_V^{(3)}$, needs to be contracted with a linear fluctuation. These two quantities behave differently under parity and thus the expectation value of their product vanishes. 
In particular this means that, in a one-loop calculation, we do not need to add any additional counterterm for $\d_z\pi_z(\vec x,t)=\rho_b\,\d_z[(1+\delta(\vec x,t)) v_z(\vec x,t)]$. We can rewrite (\ref{eq:delta_redshift_2}) as
\bea\label{eq:delta_redshift_3}
\delta_r(\vec k) \simeq \delta(\vec k)+\frac{k_z^2}{k^2}\;\frac{\dot\delta(\vec k)}{H}-\epsilon^{zij}\frac{k_zk_i}{k^2}\frac{\pi_{V,j}}{H}+\frac{i^2}{2}\left(\frac{k_z}{aH}\right)^2 [v_z^2]_\vk-\frac{i^3}{3!}\left(\frac{k_z}{aH}\right)^3 [v_z^3]_\vk+\frac{i^2}{2}\left(\frac{k_z}{aH}\right)^2 [v_z^2\delta]_\vk. \nn \\
\eea
Even after we have renormalized the velocity field, the operators $[v_z^2]_\vk,\; [v_z^3]_\vk$ and $[v_z^2\delta]_\vk$ need to be renormalized. We therefore define the renormalized operators by adding suitable counterterms. The equivalence principle implies that the counterterms can only be a function of $\d_i\d_j\phi$ and $\d_i v_j$ or time derivatives along the flow. Since in this paper we will perform the calculation only at one loop order and it will suffice to evaluate the counterterms only at linear level, we can use the linear equations of motion to simplify the form of the counterterms. We obtain:
\bea\label{eq:composite_renormalization_1}
&&[v_z^2]_{R,\vk}=\hat z^i \hat z^j \left\{[v_i v_j]_{\vk}+\left(\frac{a H}{\knl}\right)^2\left[c_1 \delta^{ij}+\left(c_2 \delta^{ij}+c_3 \frac{k^i k^j}{k^2}\right) \delta(\vk)\right]+\ldots \right\} \\ \nonumber
&&\quad\qquad= [v_z^2]_\vk+\left(\frac{a H}{\knl}\right)^2 \left[c_1 +c_2\,\delta(\vk) \right]+\left(\frac{a H}{\knl}\right)^2 c_3\, \frac{k_z^2}{k^2}\, \delta(\vk)+\ldots\ , 
\\ \nonumber
&&[v_z^3]_{R,\vk}=\hat z^i \hat z^j \hat z^l \left\{[v_i v_j v_l]_{k}+\left(\frac{a H}{\knl}\right)^2 c_1 \left(\delta_{ij}\,v_l(\vk)+\text{2 permutations}\right)+\ldots   \right\} \\ \nonumber
&&\quad\qquad= [v_z^3]_\vk+\left(\frac{a H}{\knl}\right)^2 3\, c_1 \, v_z(\vk)+\ldots \ , \\ \nonumber
\eea
\bea \nonumber
&&[v_z^2\delta]_{R,\vk}=\hat z^i \hat z^j  \left\{[v_i v_j \delta]_{k}+\left(\frac{a H}{\knl}\right)^2 c_1 \; \delta_{ij}\,\delta(\vk) +\ldots \right\} \\ \nonumber
&&\quad\qquad= [v_z^2 \delta]_\vk+\left(\frac{a H}{\knl}\right)^2 \, c_1 \, \delta(\vk)+\ldots \ .
\eea
For the first of the operator above, $[v_z^2]_\vk$, we have renormalized both the vacuum expectation value and the correlation with external long wavelength fields. In the second and third operators, only the vacuum expectation value is important at one loop order. In equation (\ref{eq:composite_renormalization_1}) the dots refer to higher derivative or higher order terms, as well as stochastic counterterms, that are irrelevant for a one-loop calculation~\footnote{It is maybe worth to spend a couple of words on the stochastic counterterms. For example, to the operator~$[v^i v^j]$, one could add a stochastic counterterm of the form  $\Delta [v^i v^j]$, with a two point function of order 
\be
\langle\left[\Delta [v^i v^j]\right]_{\vec k}\left[\Delta [v^i v^j]\right]_{\vec k'}\rangle\sim\delta^{(3)}(\vec k +\vec k') \left(\frac{H^2}{\knl^2}\right)^2\frac{1}{\knl^3}\ .
\ee It is now straightforward to realizes that this terms contributes to the matter power spectrum as $\left(k/\knl\right)^4$, as it therefore negligible at one loop order. }. The factors of $aH/\knl$ have been chosen so that the parameters $c_i$'s are expected to be order one numbers. 

Eq.~(\ref{eq:delta_redshift_3}) should therefore be meant as written in terms of the renormalized contact operators involving the velocity and the density fields:
\bea\label{eq:delta_redshift_4}
&&\delta_r(\vec k)\simeq\\ 
&&\ \   \delta(\vec k)+\frac{k_z^2}{k^2}\,\frac{\dot\delta(\vec k)}{H}-\epsilon^{zij}\frac{k_zk_i}{k^2}\frac{\pi_{V,j}}{H}+\frac{i^2}{2}\left(\frac{k_z}{aH}\right)^2 [v_z^2]_{R,\vk}-\frac{i^3}{3!}\left(\frac{k_z}{aH}\right)^3 [v_z^3]_{R,\vk}+\frac{i^2}{2}\left(\frac{k_z}{aH}\right)^2 [v_z^2\delta]_{R,\vk} . \nn  
\eea
Written in terms of the bare (that is non-renormalized) operators, we have, for $k\neq 0$,
\bea\label{eq:delta_redshift_5}
&&\delta_r(\vec k) \simeq  \delta(\vec k)+\frac{k_z^2}{k^2}\,\frac{\dot\delta(\vec k)}{H}-\epsilon^{zij}\frac{k_zk_i}{k^2}\frac{\pi_{V,j}}{H}\\ \nonumber
&&\qquad\quad +\frac{i^2}{2}\left(\frac{k_z}{aH}\right)^2\left( [v_z^2]_{\vk}+\left(\frac{a H}{\knl}\right)^2 c_2\,\delta(\vk)+\left(\frac{a H}{\knl}\right)^2 c_3\, \frac{k_z^2}{k^2}\, \delta(\vk)\right)\\ \nonumber
&&\qquad\quad-\frac{i^3}{3!}\left(\frac{k_z}{aH}\right)^3\left( [v_z^3]_{\vk}+\left(\frac{a H}{\knl}\right)^2 3\, c_1 \, v_z(\vk)\right)\\ \nonumber
&&\qquad\quad+\frac{i^2}{2}\left(\frac{k_z}{aH}\right)^2\left( [v_z^2\delta]_{\vk}+\left(\frac{a H}{\knl}\right)^2 \, c_1 \, \delta(\vk)\right)\ .
\eea
or, more simply,
\bea\label{eqm1}
&&\delta_r(\vec k)= \delta(\vec k)+\mu^2 \,\frac{\dot\delta(\vec k)}{H}-\epsilon^{zij}\frac{\mu\, k_i}{k}\frac{\pi_{V,j}}{H}-\frac{1}{2}\left(\frac{k\, \mu}{a H}\right)^2 \left([v_z^2]_\vk+[v_z^2\delta]_\vk\right)+\frac{i}{6}\left(\frac{k\, \mu}{a H}\right)^3 [v_z^3]_\vk\\ \nonumber
&&\qquad\quad+\frac{c_1}{2}\mu^4\left(\frac{k}{\knl}\right)^2  \frac{\theta(\vk)}{aH}
-\frac{(c_1+c_2)}{2}\mu^2 \left(\frac{k}{\knl}\right)^2 \delta(\vk)-\frac{c_3}{2}\mu^4 \left(\frac{k}{\knl}\right)^2 \delta(\vk)\ ,
\eea
where we can use the fact that in the EFTofLSS the curl of the velocity field appears  only at very high order in perturbation theory~\cite{Carrasco:2013mua}:  for a one-loop calculation we can therefore write $v_i(\vec x,t)=\frac{\d_i}{\d^2}\theta(\vec x,t)$. We also defined $\mu=\hat k\cdot\hat z$.
The terms in the first line of equation (\ref{eqm1}) are the standard terms that appear in SPT (see for example Appendix B of~\cite{Matsubara:2007wj}). However, there is an important difference with respect to SPT: the equations of motion that are obeyed by the $\delta$ and $\theta$ fields are {\it not} the SPT ones but rather the ones of the EFTofLSS, with their own counterterms. The terms in the second line  of equation (\ref{eqm1}) are new. They are the contribution of the counterterms that only affect the mapping to redshift space. At one-loop order, these terms will be evaluated at linear order.

The formula for the power spectrum in redshift space therefore reads
\bea
&&P_{r,\delta,\delta,\,||_\text{1-loop}}(k,\mu,t)=P_{\delta,\delta,||_\text{1-loop}}(k,t)+2\mu^2 P_{\delta,\frac{\dot\delta}{H},||_\text{1-loop}}(k,t)\\ \nonumber 
&&\qquad\qquad\qquad+\mu^4P_{\frac{\dot\delta}{H},\frac{\dot\delta}{H},||_\text{1-loop}}(k,t)-\left(\frac{k\, \mu}{aH}\right)^2P_{\delta,[v_z^2],\text{tree}}(k,t)\\ \nonumber
&&\qquad\qquad\qquad-\mu^2\left(\frac{k\, \mu}{aH}\right)^2  P_{\frac{\dot\delta}{H},[v_z^2],\text{tree}}(k,t)+\frac{1}{4} \left(\frac{k\, \mu}{aH}\right)^4 P_{[v_z^2],[v_z^2],\text{tree}}(k,t)\\ \nonumber
&&\qquad\qquad\qquad+\left(1+f\mu^2\right)\left(\frac{k\, \mu}{aH}\right)^2 P_{\delta,[\delta\,v_z^2],\text{tree}}(k,t)+\frac{i}{3}\left(1+f\mu^2\right)\left(\frac{k\, \mu}{aH}\right)^2 P_{\delta,[v_z^3],\text{tree}}(k,t)\\ \nonumber
&&\qquad\qquad\qquad-\left(1+f \mu^2\right) \left[\left(c_1+c_2\right) \mu^2+\left( c_1+c_3\right) \mu^4\right]\left(\frac{k}{\knl}\right)^2P_{\delta,\delta,11}(k,t)\ ,
\eea
where we have defined the logarithmic derivative of the growth factor $D$ as $f=\frac{\d\log D}{d\log a}$ and in the third line we have used that 
$P_{\frac{\dot\delta}{H},[\delta\,v_z^2],\text{tree}}(k,t)=f\, P_{\delta,[\delta\,v_z^2],\text{tree}}(k,t)$ and similarly that $P_{\frac{\dot\delta}{H},[v_z^3](k,t),\text{tree}}=f P_{\delta,[v_z^3](k,t),\text{tree}}$. The notation ${}_{||_\text{$N$-loops}}$ means that we are evaluating a quantity {\it up to} order $N$ loops. In the last line, we have re-absorbed a factor of $f$ inside the counterterm $c_1$, which is time dependent anyway: $f c_1\to c_1$. 

The terms in the first four lines represent terms that need to be computed in the EFTofLSS in real space. $P_{\delta,\delta,\text{1-loop}},\;P_{\delta,\frac{\dot\delta}{H},\text{1-loop}},\;$ and $P_{\frac{\dot\delta}{H},\frac{\dot\delta}{H},\text{1-loop}}$ have been already computed in~\cite{Carrasco:2012cv,Senatore:2014via}. The others can be computed in a similar way. One can use the SPT formulas of Appendix B of~\cite{Matsubara:2007wj}, with the addition of the relevant counterterms that we explain next. 

Instead the terms on the last line are the contribution from the counterterms that appear directly only in redshift space. Alternatively, these counterterms can thought of as appearing in the EFTofLSS calculation of correlation functions involving quadratic operators such as $P_{\delta,[v_z^2],\text{tree}}(k,t)$. We see that, at each redshift, while naively there are three new counterterms, at one-loop one linear combination is degenerate, so that we need only two counterterms. The two independent combinations scale as respectively as $(1+f\mu^2)\mu^2 k^2 P_{\delta,\delta,11}$ and $(1+f\mu^2)\mu^4 k^2 P_{\delta,\delta,11}$. Notice that there is no independent counterterm for each $\mu$-dependence that appears in the SPT formulas.

In terms of the familiar SPT contributions, we can write the above expression as
\bea
&&P_{r,\delta,\delta,||_\text{1-loop}}(k,\mu,a)= D^2\,P_{r,\delta,\delta,11}(k,\mu)+D^4\, P_{r,\delta,\delta,\text{1-loop,\,SPT}}(k,\mu)\\ \nonumber
&&\qquad\qquad -(2\pi)D^4 \left[2 c_s^2 +\mu^2\left(8 \, c_s^2\,f +2\frac{d\, c_s^2}{d\log a}\,\right)  +2\, \mu^4    \left(3 c_s^2\, f^2+\frac{d\, c_s^2}{d\log a}\, f \right)\right. \\ \nonumber
&&\qquad\qquad\qquad\qquad\qquad\qquad \left.- \left(1+f \mu^2\right) \left(\left(c_1+c_2\right) \mu^2+\left( c_1+c_3\right) \mu^4\right)\right]\left(\frac{k}{\knl}\right)^2P_{\delta,\delta, 11}(k) \ ,
\eea
where an explicit expression for $P_{r,\delta,\delta,\text{1-loop, SPT}}(k,\mu)$ is given in Appendix~\ref{app:IR-safety}. We have redefined the parameters $c_{1,2,3}$ by extracting a factor of $(2\pi)D^2$: $c_{i}\to(2\pi) D^2 c_i$, and we have used the linear power spectrum of dark matter at present time. All $c_i$'s are now expected to be order one. We have also written the counterterm for the dark matter power spectrum as $P_{\delta\delta,\text{counter}}(k,t)=-2(2\pi)c_s^2 D^4 \frac{k^2}{\knl^2} P_{\delta,\delta,11}(k)$. With these conventions, in a scaling universe with slope of the power spectrum equal to $n$, the finite part of $c_s^2$ would have a time dependence $\propto D^{\frac{4}{3+n}-2}$, while the part proportional to the cutoff of the loop integrals is time-independent. 

The matter power spectrum is IR safe not only in real space, but also in redshift space. The reason is very similar to the one that makes the dark matter IR-safe in real space~\cite{Carrasco:2013sva}. The IR divergencies are due to the contribution of the long wavelength displacements, which, by the equations of motion, are proportional to gradients of the Newtonian potential. In General Relativity the role of the Newtonian potential is simply kinematical, needed to identify the position of an object on a manifold. It does not have any dynamical effect as it can be set to zero by a coordinate transformation ({\it ie.} going to the local Fermi frame). Therefore the effect of the displacements is irrelevant for equal time matter correlation functions and thus real space IR divergencies are absent for equal time correlation functions. 

This General Relativistic argument has been formulated in~\cite{Carrasco:2013sva} and explains in a general relativistic setting the previous explanations of~\cite{Scoccimarro:1995if}. It has subsequently been used in~\cite{Creminelli:2013mca} to explain in a general relativistic context the so-called consistency conditions in large scale structures~\cite{Riotto,Peloso:2013zw,Kehagias:2013rpa}. 

The difference between real space and redshift space is due to the velocity, which is  given by the time derivative of the displacement and so the same logic applies: equal time matter correlation functions are IR safe in redshift space. As it was originally explained in~\cite{Senatore:2014via} and we repeat next, this argument applies only to the extreme infrared modes. In the current universe there are effects from intermediate modes that give infrared effects that do not cancel and need to be treated in a non-perturbative way.  

We have used the formulas of~\cite{Matsubara:2007wj}, supplemented with the additional terms from the EFTofLSS, to explicitly check the cancellation of IR divergencies. Notice that many individual terms, such as $P_{\frac{\dot\delta}{H},\frac{\dot\delta}{H},\text{1-loop}}$ and $P_{\frac{\dot\delta}{H},[v_z^2],\text{tree}}(k)$ are not IR-safe when taken on their own. The cancellation happens within the sum of all terms. In the same spirit as~\cite{Carrasco:2013sva}, in appendix~\ref{app:IR-safety}, we provide an expression  for  the loop integrand that is directly IR-safe, a fact that is convenient for numerical computations.  In appendix~\ref{app:UV-renormalization} we show that the counterterms we add in the EFTofLSS are necessary to renormalize and make finite the predictions for scaling universes, which offers a nice consistency check of our formulas~\footnote{
It should be emphasized that the fact that the calculation of redshift space distortions is sensitive to the contribution of short distance fluctuations, and therefore not correctly described by the standard SPT formulas has been already noticed in~\cite{Vlah:2012ni,Vlah:2013lia}. The way these references address the issue is at the same time similar and different from our approach. Restricting themselves to an Eulerian calculation, they indeed identify the terms that are UV sensitive and add counterterms. In detail, they identify the terms in the final result that contain the velocity dispersion at a given location, and adjust them by adding an additional contribution. They use a different correction for each ocurrence of the velocity dispersion. The number and the  functional form of the resulting counterterms is in general different than ours. It is also not very clear what the perturbative order one assigns to these new terms. Thus it does not appear straightforward to  generalize the scheme to higher orders (as an example of this kind of subtleties one could look at what happens in the  EFTofLSS applied to dark matter at two-loop order~\cite{Carrasco:2013mua}). It is also not obvious  how to impose the fact that  the allowed counterterms are not-always linearly  independent.  All of these issues are instead automatically and somewhat trivially addressed in our approach because the inclusion of the counterterms is done by adding suitable fields, not just parameters. These fields are solved for following the same perturbative scheme that is used to solve for all fields.  Their importance is explicitly dictated by an expansion in powers of $\sim k/\knl$. The usual renormalization techniques, as the ones described in~\cite{Carrasco:2012cv,Pajer:2013jj,Carrasco:2013mua}, can be straightforwardly applied. 
}.

\subsubsection{Biased tracers}

The formalism we have developed so far for dark matter can be easily extended to biased tracers. The dynamics of halos or galaxies, to which we will sometime refer to as collapsed objects, in real space was discussed in the context of the EFTofLSS in~\cite{Senatore:2014eva}. We refer to~\cite{Senatore:2014eva}  for the main results and the notation. The relationship between the overdensity of tracers in redshift space, $\delta_{M,r}$, and the one in real space, $\delta_{M}$, is given by the same formula (\ref{eq:first}), with obvious change of labels:
\be\label{eq:first_halo}
\delta_{M,r}(\vec k)=\delta_M(\vec k)+\int d^3 x \; e^{-i \vec k\cdot\vec x} \left(\exp\left[-i \frac{k_z }{a H} v_{M,z}(\vec x)\right]-1\right)\left(1+\delta_M(\vec x)\right)\ ,
\ee
where $\vec v_{M}(\vec x,t)$ is the real space velocity of halos. By Taylor expansion, we arrive at the analogous of (\ref{eq:delta_redshift_1}):
\bea\label{eq:delta_redshift_2_halos}
&&\delta_{M,r}(\vec k) =\delta_M(\vec k)-i \frac{k_z}{aH} v_{M,z}(\vec k)\\ \nonumber
&&\qquad\qquad-i \frac{k_z}{aH} [\delta_M v_{M,z}]_\vk+\frac{i^2}{2}\left(\frac{k_z}{aH}\right)^2 [v_{M,z}^2]_\vk-\frac{i^3}{3!}\left(\frac{k_z}{a H}\right)^3 [v_{M,z}^3]_\vk+\frac{i^2}{2}\left(\frac{k_z}{aH}\right)^2 [v_{M,z}^2\delta]_\vk . \nonumber 
\eea
In the case of halos we do not replace the velocity field with the momentum field. This replacement is less useful for collapsed objects because matter and momentum conservation laws do not apply to these. Relative  to the case of dark matter we have one additional contact operator to renormalize:~$[\delta_M v_{M,z}]$. In analogy to (\ref{eq:composite_renormalization_1}), we can write
\bea\label{eq:composite_renormalization_2}\nonumber
&&[v_{M,z}^2]_{R,\vk}=\hat z^i \hat z^j \left\{[v_{M,i} v_{M,j}]_{\vk}+\left(\frac{a H}{\km}\right)^2\left[c_{1,M} \delta^{ij}+\left(c_{2,M} \delta^{ij}+c_{3,M} \frac{k^i k^j}{k^2}\right) \delta(\vk)\right]+\ldots \right\}  \\ \nonumber
&&\qquad\qquad= [v_{M,z}^2]_\vk+\left(\frac{a H}{\km}\right)^2 \left[c_{1,M} +c_{2,M}\,\delta(\vk) \right]+\left(\frac{a H}{\km}\right)^2 c_{3,M}\, \frac{k_z^2}{k^2}\, \delta(\vk)+\ldots\ ,
\\  \ \nonumber
&&[v_{M,z}^3]_{R,\vk}=\hat z^i \hat z^j \hat z^l \left\{[v_{M,i} v_{M,j} v_{M,l}]_{k}+\left(\frac{a H}{\km}\right)^2 c_{1,M} \left(\delta_{ij}\,v_{M,l}(\vk)+\text{2 permutations}\right)+\ldots   \right\} \\ \nonumber
&&\qquad\qquad= [v_{M,z}^3]_\vk+\left(\frac{a H}{\km}\right)^2 3\, c_{1,M} \, v_{M,z}(\vk)+\ldots \ ,\\ \nonumber
&&[v_{M,z}\delta]_{R,\vk}=\hat z^i  \left\{[v_{M,i}  \delta]_{k}+\left(\frac{a H}{\km}\right)i\, c_{4,M} \; \frac{k_i}{\km}\,\delta(\vk) +\ldots \right\} \\ \nonumber
&&\qquad\qquad= [v_{M,z} \delta]_\vk+\left(\frac{a H}{\km}\right) \, c_{4,M} \frac{k_z}{\km}\, \, \delta(\vk)+\ldots \ ,\\ \nonumber
&&[v_{M,z}^2\delta]_{R,\vk}=\hat z^i \hat z^j  \left\{[v_{M,i} v_{M,j} \delta]_{k}+\left(\frac{a H}{\km}\right)^2 c_{1,M} \; \delta_{ij}\,\delta(\vk) +\ldots \right\} \\ 
&&\qquad\qquad= [v_{M,z}^2 \delta]_\vk+\left(\frac{a H}{\km}\right)^2 \, c_{1,M} \, \delta(\vk)+\ldots \ .
\eea
There are a few differences to point out in this set of equations with respect to the ones that we wrote for dark matter in~(\ref{eq:composite_renormalization_1}). First, we expressed the response of the  collapsed-objects composite operators directly in terms of the dark matter fields. This is useful because, at the end of day, we compute real-space correlation functions of collapsed objects through bias coefficients times correlation functions of  dark matter fields. Therefore, it is useful to skip a step and write the response directly in terms of dark matter fields. This is why the derivative expansion is controlled by a new parameter $\km$, rather than $\knl$. The two parameters might be different and $\km$ is expected to depend on the mass of the collapsed object~\cite{Senatore:2014eva}.  As in the case of dark matter, the stochastic counterterms are irrelevant at one loop and have therefore been neglected.

As a result of the small difference in the structure of the counter terms, the formula for the power spectrum in redshift space now reads:
\bea\nonumber
&&P_{r,\delta_M,\delta_M,||_\text{1-loop}}(k)=P_{\delta_M,\delta_M,||_\text{1-loop}}(k)-2 i \left(\frac{k\,\mu}{aH}\right) P_{\delta_M,v_{M,z},||_\text{1-loop}}(k)-\left(\frac{k\,\mu}{aH}\right)^2 P_{v_{M,z},v_{M,z},||_\text{1-loop}}(k)\\ \nonumber
&&\qquad-2 i\left(\frac{k\, \mu}{aH}\right)P_{\delta_M,[\delta_M v_{M,z}],\text{tree}}(k)-2 \left(\frac{k\, \mu}{aH}\right)^2P_{v_{M,z},[\delta_Mv_{M,z}],\text{tree}}(k)\\ \nonumber
&&\qquad-\left(\frac{k\, \mu}{aH}\right)^2P_{\delta_M,[v_{M,z}^2],\text{tree}}(k)+i \left(\frac{k\, \mu}{aH}\right)^3P_{v_{M,z},[v_{M,z}^2],\text{tree}}(k)\\ \nonumber
&&\qquad+ \frac{i}{3}\left(\frac{k\, \mu}{aH}\right)^3P_{\delta_M,[v_{M,z}^3],\text{tree}}(k)+ \left(\frac{k\, \mu}{aH}\right)^4 P_{v_{M,z},[v_{M,z}^3],\text{tree}}(k)\\ \nonumber
&&\qquad- \left(\frac{k\, \mu}{aH}\right)P_{\delta_M,[\delta_M v_{M,z}^2],\text{tree}}(k)+i \left(\frac{k\, \mu}{aH}\right)^3P_{v_{M,z},[\delta_Mv_{M,z}^2],\text{tree}}(k)\\ \nonumber
&&\qquad- (2\pi)\left[\left(c_{1,M}+c_{2,M}-2c_{4,M}\right) \mu^2+\left( c_{1,M}+c_{3,M}\right) \mu^4\right]\left(\frac{k}{\km}\right)^2P_{\delta_M,\delta,\text{tree}}(k)\\ 
&&\qquad- (2\pi)\left[\left(c_{1,M}+c_{2,M}-2c_{4,M}\right) \mu^2+\left( c_{1,M}+c_{3,M}\right) \mu^4\right] \left(\frac{k}{\km}\right)^2\left(\frac{k\, \mu}{aH}\right) P_{v_{M,z},\delta,\text{tree}}(k) \ .
\eea
Notice that the term $\left(\frac{k\, \mu}{aH}\right) P_{v_{M,z},\delta,\text{tree}}(k)$ is expected to be comparable in size to  $ P_{\delta_M,\delta,\text{tree}}(k)$. As for dark matter, we have redefined the counterterm by extracting a factor of $(2\pi)$.

\subsection{IR-resummation\label{sec:IR-resummation}}

In the former subsection we have obtained expressions for the power spectrum of matter in redshift space by expanding in $\epssm$, $\epssM$ and $\epsdm$ and treating them on equal footing. As it was discussed in detail in~\cite{Senatore:2014via}, this treatment is not sufficient to reproduce the features of the BAO oscillations at low redshifts as well as to correctly compute non-IR-safe qualities. This is due to the particular initial conditions in our universe that are characterized to large coherent velocities and  displacements of the dark matter particles. At redshift zero the typical size of these displacement is of the order of the width of the BAO peak in the correlation function. This  means that it cannot be accurately  reconstructed using a small number of terms in a Taylor expansion.  

A treatment where we do not Taylor expand in the long wavelength displacements $\epssm$, so called IR-resummation, was developed in~\cite{Senatore:2014via} for quantities in real space. We showed how by doing this the BAO peak could be better reconstructed and non-IR-safe quantities, such as the momentum power spectrum, were also better reproduced. 

When discussing quantities in redshift space, the displacement field is not the only quantity that needs to be treated non-perturbatively. The velocity of a certain region affects the redshift-space position of a particle proportionally to $k_z v_z(\vec x)/H$ (for example in the exponential of~(\ref{eq:first})). When $k$ is a short mode such that  $k_z v_z(\vec x)/H \gg1$ one needs to go to high order in a Taylor expansion for it to start to converge. This is indeed the same parameter that controls the effect  in real space because displacements are of order $v/H$.
 
To describe the IR-resummation it is more useful to work in a Lagrangian description.  The Lagrangian description in the context of the EFTofLSS was developed in~\cite{Porto:2013qua}. There it is shown that the universe can be thought of as being made up of {\it extended} particles that are therefore identified not only by their center of mass position $\vec z(\vec q,t)$, with $\vec q$ being a label that identifies a particle. One also needs to give  the properties associated to the extended nature of the particles, such as quadrupole, octupole and higher multipole moments. The properties associated to their finite size are important to correctly reproduce the effect that strongly-coupled short distance physics has on large distances. 

We are going to use this Lagrangian point of view simply to resum the long wavelength displacement fields. As we will see, we will be able to do this by reducing the calculation to manipulation of the Eulerian calculation. For the purpose of keeping the formulas as simple as possible we will not explicitly write down the contributions from the quadruple, octupole, etc of each particle and refer the interested reader to~\cite{Porto:2013qua}.  Our construction indeed resembles very closely the one we did in~\cite{Senatore:2014via} for the real space EFTofLSS as applied to dark matter, and in~\cite{Senatore:2014eva} for the EFTofLSS as applied to biased objects.

With these caveats, in real space the overdensity of particles is given by
\be
\delta(\vec x,t)=\int d^3q\; \delta^{(3)}(\vec x-\vec z(\vec q,t))=\int d^3q\;\int \frac{d^3k}{(2\pi)^3}\; e^{- i\,\vec k\cdot\left(\vec q+\vec s(\vec q,t)\right)}\ ,
\ee
where we defined the displacement field $\vec s$ as $\vec z(\vec q,t)=\vec q+\vec s(\vec q,t)$. The Eulerian calculation corresponds to Taylor expanding in $\vec s$, while we wish to resum all factor of $\vec s$ that appear without a gradient acting upon them. 

In redshift space, the formula above simply generalizes to
\bea
&&\delta_r(\vec x_r,t)=\int d^3q\; \delta^{(2)}\left(\xperp-\zperp\right) \;\delta^{(1)}\left( \xparal- \zparal-\frac{\dsparal}{H}\right)\\ \nonumber
&&\qquad\qquad=\int d^2q_\perp dq_\parallel\;\int \frac{d^2k_\perp}{(2\pi)^2} \frac{dk_\parallel}{(2\pi)}\; e^{- i\,\kperp\cdot\left(\qperp+\sperp\right)}\;e^{- i\,\kparal\cdot\left(\qparal+\sparal+\frac{\dsparal}{H}\right)}\ .
\eea
Since for modes where non-linear corrections are important evolution effects can be neglected, we can go to 3-dimensional Fourier space:
\be
\delta_r(\kperp,\kparal,t)=\int d^2q_\perp dq_\parallel\; e^{- i\,\kperp\cdot\left(\qperp+\sperp\right)}\;e^{- i\,\kparal\cdot\left(\qparal+\sparal+\frac{\dsparal}{H}\right)}\ .
\ee
The power spectrum in redshift space can then be obtained by taking the expectation value of $\langle\delta_r(\kperp,\kparal,t)\delta_r(\kperp',\kparal',t')\rangle$ and performing the integral with respect to the average Lagrangian coordinates. This integration leads to a factor of $\delta^{(2)}(\kperp+\kperp')\delta^{(1)}(\kparal+\kparal')$, as the integral does not depend on the overall position in $q$-space. Since we are neglecting evolution effects, we will focus only on the equal time correlation functions. All of this allows us to write the power spectrum as
\bea
&&P_{\delta,\delta,r}(k_\perp,\kparal,t)\\ \nonumber
&&\qquad\qquad =\int d^2q_\perp dq_\parallel\;  e^{-i\, \kperp\cdot\qperp-i\, \kparal\cdot\qparal}\left\langle e^{-i\, \kperp\cdot \left(\sperp-\sperpo\right)-i\, \kparal\cdot\left(\sparal-\sparalo+\frac{\dsparal-\dsparalo}{H}\right)}\right\rangle\ .
\eea
By using the cumulant theorem, we can write
\be
P_{\delta,\delta,r}(k_\perp,\kparal,t)=\int d^2q_\perp dq_\parallel\;  e^{-i\, \kperp\cdot\qperp-i\, \kparal\cdot\qparal} \; K_r(\kperp,\kparal,\qperp,\qparal;t)\ ,
\ee
with
\bea
&&\!\!\!\!\!\!\!\!\!\!\!\! K_r(\kperp,\kparal,\qperp,\qparal;t)=\\ \nonumber
&&\!\!\!\!\!\!\!\!\!\!\!\! \exp\left[\sum_{N=0}^\infty \frac{1}{N!}\left\langle\left(-i\, \kperp\cdot \left(\sperp-\sperpo\right)-i\, \kparal\cdot\left(\sparal-\sparalo+\frac{\dsparal-\dsparalo}{H}\right)\right)^N \right\rangle\right].
\eea

The derivation at this point follows in strict analogy the one done in~\cite{Senatore:2014via}, so we will proceed very swiftly and refer to~\cite{Senatore:2014via} for details. We will also use the additional simplification of the final formulas of~\cite{Senatore:2014via} derived in Appendix C of~\cite{Angulo:2014tfa}. 

We are interested in resumming only the linear part of the displacement, so that we can define
\bea
&&\!\!\!\!\!\!\!\!\!\!\!\! K_{r,0}(\kperp,\kparal,\qperp,\qparal;t)=\\ \nonumber
&&\!\!\!\!\!\!\!\!\!\!\!\! \exp\left[- \frac{1}{2}\left\langle\left( \kperp\cdot \left(\sperp_1-\sperpo_1\right)+ \kparal\cdot\left(\sparal_1-\sparalo_1+\frac{\dsparal_1-\dsparalo_1}{H}\right)\right)^2 \right\rangle\right].
\eea
where the subscript $_1$ represents that we take only the linear solution~\footnote{This construction can be generalized to resumming non-linear displacement. See~\cite{Senatore:2014via} for details.}. This expression can be simplified further using the fact that $\dot {\vec s}_1=H f \vec s_1$, with $f=\d\log D/\d\log a$ and $D$ the growth factor~\footnote{This statement is true only by assuming that we can approximate the time dependence of the fluctuations with the one obtained in EdS after replacing factors of $a$ with factors of the growth factor $D$. This is an allowed approximation  that has been checked to be  accurate to percent level The purpose of our IR-resummation is to resum the greatest part of the displacement field parametrized by $\epssm$. Once this resummation is done with a good accuracy, the part that is not correctly resumed, parametrized by $\tilde{\epsilon}_{s<}$, is much smaller than one and is recovered order by order in perturbation theory~\cite{Senatore:2014via}. }. We can write
\bea
&& K_{r,0}(\kperp,\kparal,\qperp,\qparal;t)= \exp\left[- \frac{1}{2}\left\langle\left(k_i B_{ij} \left(\vec s_1(\vec q,t)-\vec s_1(\vec 0,t)\right)^j\right)^2 \right\rangle\right] \\ \nonumber
&&\qquad\qquad\qquad\qquad\quad=\exp\left[- \frac{1}{2}\left\langle\left(\vec{\tilde k} \cdot \left(\vec s_1(\vec q,t)-\vec s_1(\vec 0,t)\right)^j\right)^2 \right\rangle\right]\ ,
\eea
where in the first line we have introduce the matrix $B_{ij}$ that in the basis $\{\kperp,\,\kparal\}$  reads
\be
B=\left(\begin{array}{cc}
1 & 0 \\
0 & 1+f
\end{array}\right)\ ,
\ee
and in the second line we have defined, for a generic vector $\vec k$, its tilded dual $\vec{\tilde k}$ as
\be
\vec {\tilde k}_i=B_{ij} k_j \ .
\ee
We notice that $ K_{r,0}(\kperp,\kparal,\qperp,\qparal;t)$ is a function of $\kperp$ and $\kparal$ only through the vector $\vec{\tilde k}=\vec{\tilde k} (\kperp,\kparal)$, and in terms of these variables it is the same function as in real space:
\be
K_{r,0}\left(\kperp,\kparal,\qperp,\qparal;t\right)=K_{0}\left(\vec{\tilde k} (\kperp,\kparal),\qperp,\qparal;t\right)\ .
\ee
We now proceed by defining 
\be
F_{||_{N-j}}\left(\vec{\tilde k} (\kperp,\kparal),\qperp,\qparal;t\right)=K_{0}\left(\vec{\tilde k} (\kperp,\kparal),\qperp,\qparal;t\right) \cdot\left.\left.K_{0}^{-1}\left(\vec{\tilde k} (\kperp,\kparal),\qperp,\qparal;t\right)\right|\right|_{N-j} \ ,
\ee
where the subscript ${}_{||_{N}}$ means that we expand {\it up to} order $N$ both in $\epssm$, $\epssM$, and $\epsdm$ counting them on equal footing (this is the standard Eulerian expansion in powers of the power spectrum). We can therefore write the following expression for the power spectrum in redshift space at order $N$ expanding both in $\epsdm$ and $\epssM$ but with no expansion in $\epssm$ as
\bea\label{eq:momentum_resum_1}
&&\left.P_{\delta,\delta,r}(\kperp,\kparal,t)\right|_
N=\sum_{j=0}^N \int d^2q_\perp dq_\parallel\; e^{-i\, \kperp\cdot\qperp-i\, \kparal\cdot\qparal} \\ \nonumber
&&\qquad\qquad\times\;  \; F_{||_{N-j}}\left(\vec{\tilde k}(\kperp,\kparal),\qperp,\qparal;t\right)\; \times\; K_r(\kperp,\kparal,\qperp,\qparal;t)_j\ .
\eea
The subscript ${}_{|_{N}}$ means that we expand {\it up to} order $N$ both in $\epsdm$, $\epssM$, but not in $\epssm$, while the subscript $_j$ mean that we take the order $j$ term of the expression after Taylor expansion in $\epsdm$, $\epssM$ and $\epssm$. Notice in particular that the Eulerian order $j$ term for the power spectrum is given as
\be
P_{\delta,\delta,r}(\kperp,\kparal,t)_j=\int d^2q_\perp dq_\parallel\;  e^{-i\, \kperp\cdot\qperp-i\, \kparal\cdot\qparal} \; K_r(\kperp,\kparal,\qperp,\qparal;t)_j\ .
\ee

It is  useful to manipulate the above expression by multiplying by 1 written as
\be
1=\int \frac{d^3 k'}{(2\pi)^3}\; (1+f)\; \delta^{(3)}(\vec{\tilde k}'-\vec{\tilde k}) 
=\int \frac{d^3 k'}{(2\pi)^3}\;  \int d^3 q' \; (1+f) \; e^{i\,  (\vec{\tilde k}'-\vec{\tilde k})\cdot \vec q'} \ ,
\ee
where, just for clarity, $d^3k'=d^2 k_\perp'd \kparal'$ and we have used that $1+f=\det A$. By exchanging $\vec k$ with $\vec k'$ or $\vec{\tilde k}$ with $\vec{\tilde k}'$ when useful, we can write
\bea\label{eq:delta4} 
&&\left.P_{\delta,\delta,r}(\kperp,\kparal;t) \rangle\right|_{N}=\sum_{j=0}^N \int \frac{d^2 k_\perp'}{(2\pi)^2}\frac{d \kparal'}{(2\pi)}\;\  \\  \nonumber
&& \qquad\times\  \left[ (1+f)\;\int   d^2 q_\perp'\;d\qparal' \; e^{i\cdot(\vec{\tilde k}'- \vec{\tilde k}) \cdot \vec q'}\;   F_{||_{N-j}}\left(\vec{\tilde k}(\kperp,\kparal),\qperp,\qparal;t\right)\right]\\ \nonumber
&&\qquad \times\ \left[ \int   d^2 q_\perp\;d\qparal \; e^{- i \kperp' \cdot \qperp}\; e^{- i \kparal' \cdot \qparal}\; K_r(\kperp',\kparal',\qperp,\qparal;t)_{j}\right]\ .
\eea
Now, as done in~\cite{Senatore:2014via}, we  replace 
\be
F_{||_{N-j}}\left(\vec{\tilde k}(\kperp,\kparal),\qperp,\qparal;t\right)\quad\to\quad F_{||_{N-j}}\left(\vec{\tilde k}(\kperp,\kparal),\qperp{}',\qparal';t\right)\ ,
\ee 
as this amounts to doing a mistake proportional to the gradient of the IR displacements, which is perturbative and gets corrected order by order in perturbation theory. We therefore obtain
\be\label{eq:deltared7} 
\left.P_{\delta,\delta,r}(k_\perp,\kparal;t_1,t_2)\right|_N=\sum_{j=0}^N\int \frac{d^2 k_\perp'}{(2\pi)^2}\;\frac{d\kparal'}{2\pi}\; M_{r,||_{N-j}}\left( \kperp,\kparal, \kperp',\kparal';t\right)\; P_{\delta,\delta,r}(k_\perp',\kparal';t)_j\ ,
\ee
where
\bea\label{eq:Meq2}
&&M_{r,||_{N-j}}\left( \kperp,\kparal,\kperp',\kparal';t\right)=\\ \nonumber
&&\qquad\qquad =(1+f) \int   d^2 q_\perp\,d\qparal \  F_{||_{N-j}}\left(\vec{\tilde k}(\kperp,\kparal),\qperp{}',\qparal';t\right) \; e^{i\cdot(\vec{\tilde k}'- \vec{\tilde k}) \cdot \vec{q}'}\ .
\eea
Notice, quite remarkably, that  apart for the prefactor of $1+f$, $M_{r,||_{N-j}}\left( \kperp,\kparal,\kperp',\kparal';t\right)$, is as a function of $\vec{\tilde k}$, the {\it same} function $M_{||_{N-j}}\left(\vec{\tilde k}, \vec{\tilde k}';t\right)$ that is obtained in real space but evaluated at a linearly transformed momentum $\vec k\to \vec{\tilde k}(\vec k)$:
\be
M_{r,||_{N-j}}\left(\kperp,\kparal,\kperp',\kparal';t\right)=(1+f)\;M_{||_{N-j}}\left(\vec{\tilde k}( \kperp,\kparal), \vec{\tilde k}'(\kperp',\kparal');t\right)
\ee
 It is therefore automatically evaluated when the  real space $M_{||_{N-j}}$ has been. 
By plugging~(\ref{eq:Meq2}) into~(\ref{eq:deltared7}), we can perform the integral over the azimuthal angle in $\kperp'$, to obtain the final expression 
\be\label{eq:deltared8} 
\left.P_{\delta,\delta,r}(k_\perp,\kparal;t)\right|_N=\sum_{j=0}^N\int d k_\perp'\;d\kparal'\; \hat M_{r,||_{N-j}}( k_\perp,\kparal,\tilde k_\perp',\tkparal';t)\; P_{\delta,\delta,r}(k_\perp',\kparal';t)_j\ ,
\ee
with 
\bea
&& \hat M_{r,||_{N-j}}( k_\perp,\kparal,\tilde k_\perp',\tkparal';t)\\ \nonumber
&&\qquad\quad=(1+f)\frac{k_\perp'}{(2\pi)^2}\int   d^2 q_\perp\,d\qparal\;  e^{-i \tkperp\cdot\qperp}\; e^{i \left(\tkparal'-\tkparal\right)\cdot \qparal}\;J_0(\tilde{k}_\perp' q_\perp)  \; F_{||_{N-j}}\left(\vec{\tilde k}(\kperp,\kparal),\qperp,\qparal;t\right)\ ,
\eea
and where we used the fact that the $\left.P_{\delta,\delta,r}(k_\perp,\kparal;t_1,t_2)\right|_N$ and $P_{\delta,\delta,r}(k_\perp',\kparal';t)_j$ do not depend on the direction of $\kperp'$ to eliminate the dependence on this direction in $\hat M_{||_{N-j}}( k_\perp,\kparal, \tilde k_\perp',\tkparal';t)$. $J_n(x)$ is the Bessel function of the first kind with index $n$. Notice that the integral in the angle of $\qperp$ cannot be performed analytically. We are therefore left with $\hat M_{||_{N-j}}( k_\perp,\kparal, k_\perp',\kparal';t)$ as expressed as a three-dimensional integral in $\qperp$ and $\qparal$, with the integral in $q_\perp$ and $q_\parallel$, for each value of $k_\parallel$, $k_\perp$ and $\phi_q$, expressible as a bi-dimensional FFT. Notice how~(\ref{eq:deltared8}) is similar to the analogous formula in real space~\cite{Senatore:2014via,Angulo:2014tfa}
\be\label{eq:deltared9} 
\left.P_{\delta,\delta}(k;t_1,t_2)\right|_N=\sum_{j=0}^N\int d k'\; \hat M_{||_{N-j}}( k, k';t)\; P_{\delta,\delta}(k';t)_j\ ,
\ee
where one has to do only a one-dimensional FFT for each $k$.

Since, if we neglect evolution effects, for every realization of the universe that has velocities in one direction, there is an otherwise identical one with opposite velocities, the power spectrum can only depend on $\mu^2$, where $\mu=\hat k\cdot \hat z$.  This means that $P_{\delta,\delta,r}(k_\perp,\kparal;t)$ can be written only as a function of $\{k,\mu^2;t\}$.

For computing real-space correlation functions,  it turns out that it is more useful to perform the IR resummation directly in coordinate space. We expect this to be the same also for correlation functions in redshift space. Proceeding again in analogy to~\cite{Senatore:2014via}, we can Fourier transform~(\ref{eq:momentum_resum_1}) and replace the argument of $F_{r,||_{N-j}}(\ldots,\vec q,\ldots)\to F_{r,||_{N-j}}(\ldots,\vec q',\ldots)$ to write the following expression for the coordinate space correlation function in redshfit space $\xi_{r,\delta,\delta}$ as:
\bea\label{eq:delta6} 
&&\left.\xi_{r,\delta,\delta}(\vec r;t)\right|_N=\sum_{j=0}^N \int d^3 q \; P_{r,||_{N-j}}(\vec r|\vec q;t) \;  \xi_{r,j}(\vec q)   \ ,  
\eea 
where 
\be\label{eq:prob_redshift}
P_{r,||_{N-j}}(\vec r|\vec q;t)=\int\frac{d^3 k}{(2\pi)^3}\; e^{-i \vec k\cdot (\vec q-\vec r)}\;F_{||_{N-j}}(\vec{\tilde k}(\vec k),\vec q;t)\ ,
\ee
can be interpreted as the probability of ending up at a redshift space distance $\vec r$ after starting from a Lagrangian distance $\vec q$; and where the dependence of $\xi_{r,\delta,\delta}$ on the direction of $\vec r$ is only through the angle $\mu_r=\hat r\cdot \hat z$. The integral in $d^3k$ is a Gaussian integral that can be performed analytically, and simple manipulations allow us to express it in terms of the real space $P_{||_{N-j}}$ as
\be
P_{r,||_{N-j}}(\vec r|\vec q;t)=\frac{1}{\det B} P_{||_{N-j}}(B^{-1}{}_{ij}\,r^j|B^{-1}{}_{ij}\,q^j;t)\ .
\ee
The analytic expressions for $P_{||_{N-j}}$ can be found in Appendix  A of~\cite{Senatore:2014via}.
Unfortunately, contrary to what happens in real space, due to the lack of rotational invariance of the matrix $B$, the resulting $P_{r,||_{N-j}}$ depends on $\hat q\cdot \hat z$ and $\hat q\cdot \hat r$, which implies that  the integral over $\hat q$ cannot be done analytically. We are therefore left with a 3-dimensional integral to be done in~(\ref{eq:delta6}), which however is expected to be numerically not too challenging due to the fact that $P_{r,||_{N-j}}(\vec r|\vec q;t)$ is peaked for $\vec r\sim \vec q$ with a radius of about $10\, $Mpc and does not have rapid oscillatory features.

It is interesting to expand the real space correlation function in Legendre polynomials, by writing
\be
\xi_{\ldots}(\vec r;t)=\sum_l P_{\ell}(\mu_r) \; \xi_{\ldots,\,\ell}(r)\ , \qquad \xi_{\ldots,\,\ell}(r;t)=\frac{2\ell +1}{2}\int^1_{-1} d\mu_r\; P_{\ell}(\mu_r)\; \xi_{\ldots}(\vec r;t)\ ,
\ee
where $P_\ell(x)$ is the $\ell$-th Legendre Polinomial. We can write
\be
 \xi_{r,\,\ell}(r;t)=\sum_j\sum_{\ell'} \int dq \; q^2\; P_{r,||_{N-j}}( r| q;t){}_{\ell,\ell'} \;  \xi_{r,\,\ell',j}(q;t)\ ,
\ee
where 
\be\label{eq:llprmatrix}
P_{r,||_{N-j}}( r| q;t){}_{\ell,\ell'}= \frac{2\ell+1}{2}  \int_0^{2\pi} d\phi_q \int_{-1}^1 d\mu_r \int_{-1}^1 d\mu_q\;P_\ell(\mu_r) \; P_{r,||_{N-j}}(\vec r|\vec q;t)\; P_{\ell'}(\mu_q)\ .
\ee
This formulation is useful because it is not expected that very high $\ell$'s will be observationally relevant. Additionally, we notice that the resummed correlation function will not have the same $\mu$ dependence as the Eulerian one. However, given that the physical effect of the long wavelength flows that we are resumming is to smooth the BAO peak in the Eulerian correlation function, it is expected that the matrix $P_{r,||_{N-j}}( r| q;t){}_{\ell,\ell'}$ should not have support at high $\ell$, and one could even try to argue that it is non-vanishing only for $\ell\lesssim\ell'$.  Therefore, in this formulation one has to perform the 3-dimensional integral in (\ref{eq:llprmatrix}) only for the relevant $\ell,\,\ell',\, q$ and $r$, rather than the 3-dimensional integral in~(\ref{eq:delta6}) for every $r$ and $\mu_r$ of interest. In the real space calculation of~\cite{Senatore:2014via}, only $\ell=0$ and $\ell'=0$ were non-zero.

Finally let us point out that for biased tracers the resummation formulas go through with only minor changes. 
The matrices $\hat M_{r,||_{N-j}}( k_\perp,\kparal,\tilde k_\perp',\tkparal';t)$ and the probability distributions $P_{r,||_{N-j}}(\vec r|\vec q;t)_{l,l^\prime}$ used to resum the Eulerian power spectrum and correlation functions for the biased tracers are identical to those used for the dark matter, because any difference between the velocities of the dark matter and the tracers is higher order and thus we are not resumming it.

\section{Conclusions}

In this paper we have described how to include redshift space distortion within the formalism of the EFTofLSS both for dark matter and for biased tracers. The change of coordinates that transforms a certain quantity computed in real space to the analogous quantity computed in redshift space depends on  products of fields evaluated at the same location. This has two important consequences. First, in order to evaluate a certain quantity in redshift space quantities that did not affect the real space calculations need to be computed.  The formalism of the EFTofLSS  should be used. Second in the context of the EFTofLSS one needs to carefully define what is meant by products of the fields at a given location. This can be done by expressing a certain real-space product of long wavelength fields as a bare operator to which some counterterms, functions of the long wavelength fields, are added. The role of these counterterms is to correctly take into account to effect that the non-linear and uncontrolled short distance physics has on these quantities when evaluated at long wavelength.  

As an example, we have provided the expression for the dark matter and the halo power spectra, in the Eulerian framework.  These formulas differ from the SPT formulas because correlation functions of operators  are evaluated using the EFTofLSS (that is suitable counterterms have been added  when evaluating these quantities in real space). Examples are the effective speed of sound and the counterterms associated to the bias coefficients for tracers. Furthermore products of fields at the same location, that appear due to the change of coordinates from real space to redshift space, are supplemented by new counterterms. For the matter power spectrum at one loop, this corresponds to adding two terms proportional to $(1+f\mu^2) \mu^2 k^2 P_{\delta,\delta,11}$ and $(1+f\mu^2) \mu^4 k^2 P_{\delta,\delta,11}$. No independent term scaling as $\mu^6$ is present at one-loop order. For the tracer power spectrum, two additional terms scaling as $(1+f\mu^2) \mu^2 k^2 P_{\delta,\delta,11}$ and $(1+f\mu^2) \mu^4 k^2 k\, \mu\, P_{v_{M,z},\delta,\text{tree}}(k)$ need to be included. We have checked that, for dark matter, the addition of these terms is actually required to make the results finite in  scaling universes. We have also written the formulas to make IR-safety manifest at the level of the integrand, a fact which is both conceptually and numerically convenient. 

If we stopped at this level, the  treatment would be incomplete because implementing an Eulerian calculation corresponds to Taylor expanding in the parameters $\epssM$, $\epsdm$ and $\epssm$. In the current universe expanding in $\epssm$ leads to a premature breaking of perturbation theory,  at such low wavenumbers that the corrections from $\epssM$ and $\epsdm$ are still perturbatively small~\cite{Porto:2013qua,Senatore:2014via}. Since $\epssm$ encapsulates the effect of long wavelength displacements that are very well described by linear theory, it is possible to resum their contributions non-perturbatively~\cite{Senatore:2014via}. We have provided this resummation formulas in redshfit space. Due to the effect in redshift space of long wavelength velocity fields, which also scale as $\epssm$, the terms that need to be resummed are not only the one associated to the displacement field, but also the ones associated to the velocity field. The resulting formulas for the IR-resummation are therefore conceptually only slightly different than their counterparts in real space both for dark matter~\cite{Senatore:2014via} and for tracers~\cite{Senatore:2014eva}.  The breaking of three dimensional rotational invariance makes the numerical implementation of the resummation potentially more challenging. 

After this has been implemented the perturbative expansion in the EFTofLSS is a manifestly convergent expansion in powers of $k/\knl$ and $k/\km$. In particular for dark matter the  correlation functions in redshift space can be written in a schematic way as:
\bea\label{eq:dark_matter_expansion}\nonumber
\!\!\!\!\!\!\!\!\!\!\!\!\!\!\!\!\!\!\!\!&&P_{r,\delta,\delta}(k)\sim P_{\delta,\delta,11}(k)\times\underbrace{\left[1+ \left(\frac{k}{\knl}\right)^2+\ldots+\left(\frac{k}{\knl}\right)^D \right]}_\text{Derivative Expansion}\underbrace{\left[1+\left(\frac{k}{\knl}\right)^{(3+n)}+\ldots+\left(\frac{k}{\knl}\right)^{(3+n)L}\right]}_\text{Loop Expansion}\\ 
\!\!\!\!\!\!\!\!\!\!\!\!\!\!\!\!\!\!&&\qquad\qquad+\underbrace{ \left[\left(\frac{k}{\knl}\right)^{4}+\left(\frac{k}{\knl}\right)^{6}+\ldots\right]}_\text{Stochastic Terms}\ ,
\eea
where $n$ is the approximate slope of the power spectrum around the modes of interest, and where for simplicity we did not write the explicit dependence on $\mu^2$. A similar expression holds for biased tracers~\cite{Senatore:2014eva}.

In this paper we have content ourselves with simply developing the formalism to include redshift space distortions in the context of the EFTofLSS. We have presented the relevant formulas at one-loop order. Higher order generalization are, at least conceptually, straightforward. We have not taken the much harder challenge of comparing our predictions to numerical data, a fact which  would require among other things the numerical implementation of the IR-resummation formulas.  We plan to do this in the near future.

\subsubsection*{Acknowledgments}

We thank P. Creminelli for discussions. L.S. is supported by DOE Early Career Award DE-FG02-12ER41854 and by NSF grant PHY-1068380. M.Z. is supported in part by the NSF grants  AST-0907969,  PHY-1213563 and AST-1409709.

\begin{appendix}

\section{\label{app:IR-safety} IR Cancellation and IR-safe Integrand}

In this appendix we provide the formulas that allow us to compute at one-loop order the matter power spectrum in redshift space. In the EFTofLSS, at a given order the hardest calculation is the one associated to the loop diagrams that appear in SPT. Therefore, most of the formulas have already appeared in the SPT literature (see Appendix~B of~\cite{Matsubara:2007wj}). Those formulas have however the inconvenience that the IR contribution cancels only after performing the integration of the loop diagrams: the integrand has large contributions from different IR region in phase space that cancel in the final result. Following the ideas of~\cite{Carrasco:2013sva}, we manipulate the available formulas to provide an integrand that has no  large IR contributions, so that the numerical integration is made much easier.

The EFTofLSS formula at one loop reads
\bea
&&P_{r,\delta,\delta}(k,\mu,a)= D(a)^2P_{r,\delta,\delta,11}(k,\mu)+D(a)^4 P_{r,\delta,\delta,\text{1-loop, SPT}}(k,\mu)\\ \nonumber
&&\qquad\qquad + D(a)^{4}\left[P_{r,\delta,\delta,\text{matter counterterms},11}(k,\mu)+P_{r,\delta,\delta,\text{redshift counterterms},11}(k,\mu)\right]\ .
\eea
We have
\bea
&&P_{r,\delta,\delta,11}(k,\mu)=(1+f\,\mu^2 )P_{\delta,\delta,11}(k)\ , 
\eea
\bea 
&&  P_{r,\delta,\delta,\text{1-loop, SPT}}(k,\mu)=\sum_{m,n}\mu^{2n} f^m \frac{k^3}{4\pi^2} \int_0^{\Lambda/k} dr \int_{-1}^1 dx\; P_{\delta,\delta,11}(k r)\\ \nonumber
&&\quad \left\{(1+f \mu^2) P_{\delta,\delta,11}(k)\, \frac{ B_{nm}(r)}{2} \right.\\ \nonumber
&&\quad+2\left.\left[P_{\delta,\delta,11}\left(k\left(1+ r^2-2 r x\right)^{1/2}\right)\frac{A_{nm}(r,x)}{2 \left(1+ r^2-2 r x\right)^{1/2}} \right]\Theta\left(\sqrt{1+r^2-2\, r\, x}-r \right)\right.\ , \\ \nonumber
&&\quad P_{r,\delta,\delta,\text{matter counterterms},11}(k,\mu)=-2\pi\left(2 c_s^2 +\mu^2\left(8 \, c_s^2\,f +2(c_s^2)'\,\right)  +2\, \mu^4    \left(3 c_s^2\, f^2+(c_s^2)'\, f \right)\right) \\ 
&&\qquad\qquad\qquad\qquad\qquad\qquad\qquad \times\; \frac{k^2}{\knl^2} P_{\delta,\delta,11}(k)\ , \\ 
&&\quad P_{r,\delta,\delta,\text{redshift counterterms},11}(k,\mu)=-2\pi\left(1+f \mu^2\right) \left(c_a\, \mu^2+c_b \, \mu^4\right)\left(\frac{k}{\knl}\right)^2P_{\delta,\delta,11}(k)\ ,
\eea 
where $P_{\delta,\delta,11}(k)$ is the linear matter power spectrum. For the one loop term, we have that the non-vanishing components of $A_{nm}$ and $B_{nm}$ are given by~\cite{Matsubara:2007wj}
\bea
&&A_{00}(r,x)=\frac{1}{98}\left(3 r +7 x- 10 r x^2\right)^2\ , \\ \nonumber
&&A_{11}(r,x)=4 A_{00}(r,x)\ ,\\ \nonumber
&&A_{12}(r,x)=\frac{1}{28}\left(1-x^2\right)\left(7-6 r^2-42 r x+48 r^2 x^2\right)\ , \\ \nonumber
&&A_{22}(r,x)=\frac{1}{196}\left[-49 +637 x^2 +42 rx \left(17 -45 x^2\right)+6 r^2\left(19-157 x^2+236 x^4\right)\right] \ , \\ \nonumber
&&A_{23}(r,x)=\frac{1}{14}\left(1-x^2\right)\left(7-42 r x-6 r^2+48 r^2 x^2 \right) \ , \\ \nonumber
&&A_{24}(r,x)=\frac{3}{16}r^2\left(1-x^2\right)^2 \ , \\ \nonumber
&&A_{33}(r,x)=\frac{1}{14}\left(-7+35 x^2+54 r x-110 r x^3+6 r^2-66 r^2 x^2+88 r^2 x^4\right)\ , \\ \nonumber
&&A_{34}(r,x)=\frac{1}{8}\left(1-x^2\right)\left(2-3 r^2-12 r x+15 r^2 x^2\right)\ , \\ \nonumber
&&A_{44}(r,x)=\frac{1}{16}\left(-4+12 x^2+3 r^2+24 r x-30 r^2 x^2-40 r x^3+35 r^2 x^4\right)\ ,
\eea
and
\bea
&&B_{00}(r)=\frac{1}{252}\left[\frac{12}{r^2}-158+100 r^2-42 r^4+\frac{3}{r^3}\left(r^2-1\right)^3\left(7 r^2+2\right)\log\left|\frac{1+r}{1-r}\right|\right]\ , \\ \nonumber
&&B_{11}(r)=3 B_{00}(r)\ , \\ \nonumber
&&B_{12}(r)=\frac{1}{168}\left[\frac{18}{r^2}-178-66 r^2+18 r^4-\frac{9}{r^3}\left(r^2-1\right)^4\log\left|\frac{1+r}{1-r}\right|\right]\ , \\ \nonumber
&&B_{22}(r)=\frac{1}{168}\left[\frac{18}{r^2}-218+126 r^2-54 r^4+\frac{9}{r^3}\left(r^2-1\right)^3\left(3r^2+1\right)\log\left|\frac{1+r}{1-r}\right|\right]\ , \\ \nonumber
&&B_{23}(r)=-\frac{2}{3}\ .
\eea
In order to make the integrand IR safe, the usual diagrams $P_{13}$ and $P_{22}$ have been joined within one common integrand. Furthermore, one can see that what would have been the usual $P_{22}$ integrand, which is the second line of the one-loop term, has been multiplied by 2 and by $\Theta\left(|\vec k-\vec q|-\vec q\right)=\Theta\left(\sqrt{1+r^2-2 r x}-r \right)$. This corresponded to the following. If we define the $P_{22}$ diagram as
\be
P_{22}(k)=\int d^3 q\; \tilde p_{22}(\vec k,\vec q)\ ,
\ee
with
\bea
&&P_{22}(k)=\int_{|\vec k-\vec q|<q} d^3 q\; \tilde p_{22}(\vec k,\vec q)+\int_{|\vec k-\vec q|>q} d^3 q\; \tilde p_{22}(\vec k,\vec q)\\ \nonumber
&&\qquad\quad =\int_{|\vec k-\vec q|<q} d^3 q\; \tilde p_{22}(\vec k,\vec q)+\int_{|\vec k-\vec q|<q} d^3 q\; \tilde p_{22}(\vec k,\vec k-\vec q)=2\int_{|\vec k-\vec q|<q} d^3 q\; \tilde p_{22}(\vec k,\vec q)\ ,
\eea
where in the last step we have used that $ \tilde p_{22}(\vec k,\vec q)= \tilde p_{22}(\vec k,\vec k-\vec q)$. This construction is made in order to enforce that the IR divergencies are cancelled at the level of the integrand: the integrand, written in the way we did, has no IR divergencies. This manipulation is very useful for numerical porpuses, as shown in~\cite{Carrasco:2013sva} for the two loop matter power spectrum.

The last terms are instead the ones associated to the counterterms. The first one, $P_{r,\delta,\delta,\text{matter counterterms},11}$, is simply the writing in redshift space of the counterterm that is present for dark matter in real space. The second one, $P_{r,\delta,\delta,\text{redshift counterterms},11}$, in instead the one induced by the new counterterms that appear in redshift space. Notice that at one-loop order we have just two counterterms, one for $(1+f \mu^2)\mu^2$ and one for $(1+f \mu^2) \mu^4$. There is no independent term proportional to $\mu^6$, and there is no term proportional to $\mu^8$.  The overall time dependence of the counterterm can be chosen as dictated by an approximate scaling symmetry that is present in our universe if we assume matter domination and a scale free power spectrum. For the modes of interest, we can take $D^{\frac{4}{3+n}-2}$, with $n\simeq -1.7$. We used $c_a= c_1+c_2$ and $c_b= c_1+c_3$, where $c_{1,2,3}$ are defined in the main text, and, after extracting the factors of $D$ that we just discussed, they are expected to have a very small time dependence.

\section{\label{app:UV-renormalization} UV Renormalization}

It is an interesting check of our EFTofLSS formulas to show that any UV divergence can be reabsorbed by the counterterms we choose. Depending of the slope of the power spectrum, the one-loop computation will have various divergent terms, each reabsorbable by various higher derivative counterterms. We will work for scale free power spectra. The flattest slope that has a divergence at one loop is $n=-1$, that we will take:
\be
P_{\delta,\delta,11}(k)=\frac{(2\pi)^3}{\knl^3}\left(\frac{k}{\knl}\right)^{-1}\ .
\ee
Since we are interested in the divergent part, we should replace the time dependence of the counterterms we gave above with the one associated to the cutoff dependent one. For the $c_s$ term, this is $D(a)^4$.
For $n=-1$, the divergent part of the one loop diagram, cutoff with momentum $\Lambda$, reads
 \bea
&& P_{\delta,\delta,\text{1-loop}, UV}(k)=\\ \nonumber
&&\qquad\qquad -\frac{32 \pi^4}{315 }\frac{k}{\knl^4}\left(1+f \mu ^2 \right) \left(61+3 \mu ^2 f \left(61+35 \mu ^2 f^2+46 \mu ^2 f+83 f \right)\right) \log \left(\frac{\Lambda }{k}\right)\ .
 \eea
Notice that the divergence structure has some non-trivial $\mu^2$ dependence. It can be checked that the following choice of counterterms cancels the UV divergencies, as it expected to be the case:
\bea
&& c_s^2{}_{\Lambda}=-\frac{61}{315 }\log\left(\frac{k}{k_{\rm ren}}\right)\ ,\qquad c_a{}_\Lambda=-f^2 \frac{166 }{105 }\log\left(\frac{k}{k_{\rm ren}}\right)\ ,\\ \nonumber
&& c_b{}_{\Lambda}=-f^2\frac{ 2(46+35 f)}{105 }\log\left(\frac{k}{k_{\rm ren}}\right)\ ,
\eea
where $k_{\rm ren}$ is a renormalization scale. To this counterterms, a time-dependent finite part should be added to give the physical result. We stress that the fact that these terms are necessary to ensure the finiteness of the results for $n=-1$ universes is just a sufficient but non-necessary condition for the justification of the presence of these  terms in the EFTofLSS.

\end{appendix}

 \begingroup\raggedright\endgroup

\end{document}